\documentclass[preprint,showpacs,showkeys,tightenlines,10pt,prl,twocolumn]{revtex4}
\usepackage{amsfonts}
\usepackage{amsmath}
\usepackage{amssymb}
\usepackage{graphicx}
\usepackage{dcolumn}
\usepackage{bm}

\begin{document}
\preprint{T1 Process}
\title[Timing the T1 Process]{Relaxation time of the topological T1
process in a two-dimensional foam}
\author{Marc Durand}
\email{marc.durand@paris7.jussieu.fr} \affiliation{Mati\`{e}re et
Syst\`{e}mes Complexes - UMR 7057 CNRS \& Universit\'{e} Paris 7 -
Denis Diderot - case 7056, 2 Place Jussieu, 75251 Paris Cedex 05,
France}
\author{Howard A. Stone}
\email{has@deas.harvard.edu}
\affiliation{DEAS, Harvard University,
 Cambridge MA 02138, USA}
 \keywords{T1 process, relaxation, foam,
rheology, Marangoni, dilational viscosity, Gibbs elasticity}
\pacs{68.03.Cd, 68.15.+e, 83.80.Iz, 82.70.Rr}

\begin{abstract}
The elementary topological T1 process in a two-dimensional foam
corresponds to the ``flip" of one soap film with respect to the
geometrical constraints. From a mechanical point of view, this T1
process is an elementary relaxation process
 through which the entire structure of an out-of-equilibrium foam evolves. The dynamics of
this elementary relaxation process has been poorly investigated and
is generally neglected during simulations of foams. We study both
experimentally and theoretically the T1 dynamics
 in a dry two-dimensional foam. We show
that the dynamics is controlled by the surface viscoelastic
properties of the soap films (surface shear plus dilatational
viscosity, $\mu_{s}+\kappa$, and Gibbs elasticity $\epsilon$), and
is independent of the shear viscosity of the bulk liquid.  Moreover,
our approach illustrates that the dynamics of T1 relaxation process
provides a convenient tool for measuring the surface rheological
properties: we obtained $\epsilon=32\pm8$ mN/m and
$\mu_{s}+\kappa=1.3\pm0.7$ mPa.m.s for SDS, and $\epsilon=65\pm12$
mN/m and $\mu_{s}+\kappa=31\pm12$ mPa.m.s for BSA, in good agreement
with values reported in the literature.

\end{abstract}
\volumeyear{2006}
\volumenumber{number}
\issuenumber{number}
\eid{identifier}
\date[Date text]{date}
\received[Received text]{date}

\revised[Revised text]{date}

\accepted[Accepted text]{date}

\published[Published text]{date}

\maketitle

Foam rheology impacts material processing and products in many
industries and so has been the subject of continuous scientific
activity over many years \cite{Weaire,Stone}. An aqueous foam acts
macroscopically as a viscoelastic medium, whose flow depends on bulk
and surface rheological properties of the phases, which, in turn,
depend on its constitutive ingredients (surfactant, polymers,
particles), the liquid fraction, the typical bubble size, and the
shear rate. At low liquid fractions (a dry foam) bubbles have
polyhedral shapes for which local mechanical and thermodynamical
equilibria lead to Plateau's laws for the main geometric
characteristics: e.g., three films meet at each junction of a
two-dimensional foam with equal angles of $120^\circ$. Consequently,
the rheology and geometry are linked, since as the foam structure is
altered, rearrangements occur until a configuration is obtained
where Plateau's laws are satisfied.

\begin{figure}[h]
\includegraphics[width=7cm]{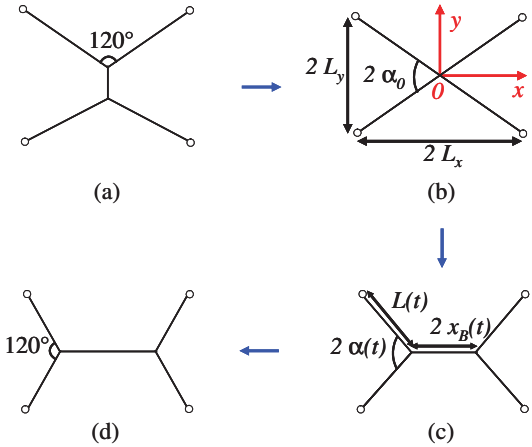}
\caption{(color online) Schematic of the T1 transition. The initial
configuration (a) evolves continuously through metastable states,
for which Plateau's laws are satisfied, to an unstable four-fold
configuration (b). This unstable state spontaneously evolves into
two three-fold junctions with creation of a new film (c) until a new
metastable configuration (d) is reached, and Plateau's laws are
satisfied again. Topologically, the transition between the initial
configuration (a) and the final configuration (b) corresponds to a
``flip" of one soap film.} \label{T1process}
\end{figure}

Any rearrangement in a two-dimensional foam may be regarded as a
combination of two elementary topological processes referred to as
T1 and T2 \cite{Weaire}. The T1 process corresponds to the ``flip"
of one soap film, as depicted in Fig. \ref{T1process}, while the T2
process corresponds to the disappearance of cells with three sides.
From a mechanical point of view, the T1 process corresponds to a
transition from one metastable configuration to another, after
passing through an unstable configuration where four films meet at
one junction (actually, for a small but finite liquid fraction, the
instability arises slightly before the four-fold vertex is formed
\cite{Hutzler}). The spontaneous evolution from one four-fold
junction to two three-fold junctions, which involves creation of a
new film, is driven by minimization of the surface area. Various
experimental and theoretical studies on the \textit{frequency} of
rearrangement events in foams have been conducted
\cite{Durian,Durian2,Earnshaw,Cohen-Addad,Marmottant}, but little is
known about the typical relaxation time associated with such events
\cite{Cox}. Indeed, the dynamics of the relaxation processes is
usually neglected in simulations of foams \cite{Aref,Kawasaki} even
though the rheological behavior of a foam obviously depends on this
relaxation time. More generally, to study the evolution of the foam
structure, it is necessary to understand the dynamics of the
elementary relaxation process.

In this paper, we investigate theoretically and experimentally the
effect of the viscoelastic parameters on the dynamics of the T1
 process. Experiments in a two-dimensional foam show that
the relaxation time depends on the interfacial viscoelasticity of
the films, but not on the shear viscosity of the bulk liquid. These
results are corroborated by a model, which allows for an estimation
of the Gibbs elasticity and the surface viscosity of the surfactants
used to make the foam.

\begin{figure}[h]
\includegraphics[width=8cm]{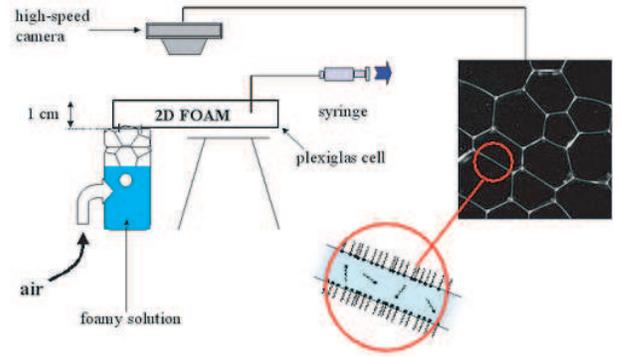}
\caption{(color online) Experimental setup: a two-dimensional foam is created in a
1 cm high horizontal Plexiglas cell by blowing air through a
surfactant solution. The  bubbles are polyhedral with an edge length
of 1-2 cm and the liquid fraction in the foam is about 1\%. Using a
syringe, air is blown away from one bubble, which induces
rearrangements, that are viewed from above using a high-speed
camera.} \label{exp_setup}
\end{figure}

The experimental setup is depicted in Fig. \ref{exp_setup}: a dry
two-dimensional foam is created in a horizontal Plexiglas cell (1 cm
high) by blowing air through a bottle containing a surfactant
solution. The polyhedral bubbles created have a typical edge length
of 1-2 cm. The liquid fraction in the foam, defined as the total
volume of liquids in the Plexiglas cell divided by the cell volume,
is about 1\%. Two different foaming agents have been used in order
to study the influence of the rheological properties of the
interface on the T1 dynamics: (i) Sodium Dodecyl Sulfate (SDS), at a
concentration of 4.80 g/L, forms ``mobile" surfaces, and (ii)
protein Bovine Serum Albumin, together with a cosurfactant Propylene
Glycol Alginate (PGA), both at concentrations of 4.00 g/L, form
``rigid" interfaces. The impact of the shear viscosity of the bulk
liquid has been investigated by adding glycerol, 0\%, 60\% and 72\%
(w/w), to the bulk solutions.

In order to cause rearrangements in the foam, we use a syringe to
 blow the air away from one bubble. Then the rearrangements are
viewed from above with a high-speed camera. The length of the soap
film is measured by following its two ends using particle-tracking
software. Experiments where the other vertices have noticeable
movements during the relaxation process, or where simultaneous T1
events occur, have been disregarded.

\begin{figure}[h]
\includegraphics[width=8cm]{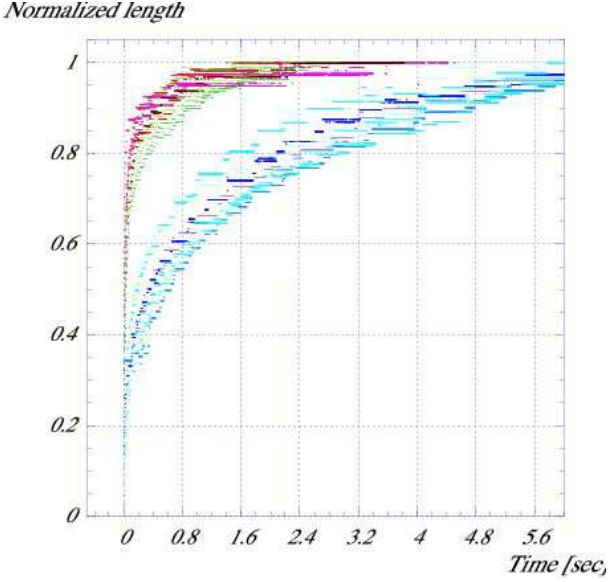}
\caption{(color online) Evolution of the film length, normalized by
its final length, with time. Each curve represents one experiment.
Red curves: SDS solution without glycerol; green curves: SDS
solution with 60\% (w/w) glycerol; blue curves: BSA/PGA solution
without glycerol. The data for SDS solutions with glycerol overlay
those without glycerol, which confirms that viscosity of the bulk
liquid is not significant. The typical time of the relaxation
process, defined as the time to reach 90\% of the final length, is
about 0.5 sec for the SDS curves and 3.7 sec for the BSA/PGA
curves.} \label{graphs}
\end{figure}

The relaxation process is characterized by the creation of a new
soap film following the appearance of a four-fold junction. For
different surfactant systems, we report
 the length of the new  film, normalized by its
final length, as a function of time in Fig. \ref{graphs}. Several
trials for each solution are shown and illustrate that the time
evolution of the reduced length appears independent of the final
film length. We define a typical time $T$ associated with the
relaxation process as the time for the film to reach 90\% of its
final length. A comparison of the results for a foam made with the
SDS without glycerol (shear viscosity of the liquid $\mu= 1.0
~\hbox{mPa.s}$) and with 60\% glycerol ($\mu= 10.7 ~\hbox{mPa.s}$)
shows that there is no significant effect of the viscosity of the
bulk liquid ($T\simeq 0.5$ sec for both solutions). This response is
not unreasonable since, for the mechanics of a free soap film,
viscous effects of the bulk are generally negligible in comparison
with the effects of the viscoelastic properties of the interfaces
\cite{Ivanov,DurandLangevin}.

We note that as the foam is constrained between two planes, most of
the liquid of each soap film is located in the menisci close to the
solid surfaces. In this region there is dissipation that depends on
the shear viscosity of the solution. Hence, from our experimental
observations we conclude that frictional effects at the boundaries
have a negligible influence on the T1 dynamics. This result is not
in contradiction with the observations made on the rheology of 2D
foam \cite{Denkov,Cantat,Cantat2}, where a macroscopic stress
(pressure drop) causes motion of the foam relative to the boundaries
and the shear viscosity of the bulk solution is the main parameter
controlling the dynamics.

Next, we compare the SDS results with those from the BSA/PGA
solution $\mu= 7~\hbox{mPa.s}$, both without glycerol. We observe a
change in the typical relaxation time by about a factor of 7
($T\simeq 3.7$ sec for the BSA/PGA solution). Although,
coincidentally, the viscosity of the BSA/PGA solution is increased
by a factor 7 relative to the SDS solution, we can rule out this
influence since we just demonstrated that the viscosity of the
solution is not rate limiting. We also verified that the addition of
glycerol to the BSA/PGA solution did not produce any significant
change in the typical relaxation times (results not shown). These
results allow us to conclude that the viscoelastic properties of the
interfaces dictate the relaxation time of the T1 process.

We now provide a brief description of a theoretical framework for
the T1 dynamics; for details of the derivation and
 complete considerations of various special cases see
\cite{DurandStoneII}. We compare the theoretical predictions with
the experiments, which allows us to extract surface rheological
parameters. The net result is a model for the T1 process and
estimates for the relaxation time.

We assume that the geometry is symmetrical with thin films nearly
together at an unstable four-fold configuration; each film connects
to one of four fixed vertices at the corner of a rectangle with
sides $2L_x$ and $2L_y$ (see Fig. \ref{T1process}). It is
experimentally observed that the stretching film has a spatially
uniform but time varying thickness $h(t)$ everywhere except near the
Plateau borders located at $x=\pm x_B(t)$ at the top and bottom
boundaries; these dynamics are common for the fluid dynamics of thin
films \cite{Breward}. We focus on the dynamics of the stretching
central film of length $2x_B(t)$, which is driven by the monotonic
decrease of the angle $\alpha (t)$:
\begin{equation}
\cos\alpha(t)=\frac{L_{x}-x_{B}(t)}{\sqrt{L_{y}^{2}+\left(  L_{x}
-x_{B}(t)\right)^{2}}}.
\end{equation}

The surface tension $\gamma$ changes in time as the surfactant
surface density $\Gamma $ is reduced by stretching. Also, we
introduce the length $L(t)=\sqrt{L_{y}^{2}+\left(
L_{x}-x_{B}(t)\right)^{2}} $ of the adjacent films which shorten
with time. Since the flat film can support no pressure gradient the
axial velocity $U(x,t)$ in the main body of the film is a linear
function of position, $U\propto x$ \cite{Breward}. This uniform
extensional motion means that away from other films we have $\gamma
(t)$ and $\Gamma (t)$, which do not change with position $x$.

The dynamics follow from a force balance on the stretching film and
a surfactant mass balance. Neglecting inertia terms and dissipative
terms associated with the viscosity of the bulk liquid \cite{DurandStoneII},
 Newton's second law applied to the film reduces to a balance between surface tension
 contributions
and surface dissipative terms ($\mu_s$ and $\kappa$ denote the shear
and dilatational viscosities and $\gamma_{eq}$ denotes the
equilibrium value of the surface tension of the soap solution):
\begin{equation}
2\gamma_{eq}\cos\alpha(t)-\gamma(t)-\left(  \mu_{s}+\kappa\right)
\frac{\partial U}{\partial x}=0.\label{equilibrium}
\end{equation}During the expansion of the new film, adjacent films act as surfactant
reservoirs, so that $\gamma(t)$ and $\Gamma(t)$ are assumed to be
close to their equilibrium values, and so are related by the
Langmuir equation of state: $\gamma(t) = \gamma_{eq} -
\epsilon\ln\left (\Gamma (t)/\Gamma_{eq}\right )$, where $\epsilon$
is the Gibbs elasticity and $\Gamma_{eq}$ is the equilibrium surface
density. This equilibrium value $\Gamma_{eq}$ is assumed to be
present in the adjacent films.

Next, we turn to a mass balance on the surfactant. This requires
accounting for stretching of the interface as well as addition of
surfactant to the {\it new} surface created as the adjacent films,
of length $L(t)$, are shortened. We assume that as surfactant is
added to the end of the stretching film the surface density is
rapidly adjusted to a spatially uniform state by strong Marangoni
forces. In addition, we neglect diffusion/adsorption processes from
the bulk liquid, restricting our study at short times (these slower
processes only affect shape adjustments as the final equilibrium is
approached). So, the change in the total number of surfactant
molecules during a time interval $dt$ is $d\left (\Gamma x_b\right
)=-\Gamma_{eq} dL$, where $dL = -dx_B.\cos\alpha (t)$; here we have
only accounted for one interface of each adjacent film as feeding
the stretched film  since surfactant from the other interface must
desorb from the surface, transit through the bulk, then adsorb to
the stretched interface, which is a much longer process. Integration
of this equation allows us to express $\Gamma(t)$ as a function of
$x_B(t)$: $\Gamma(t)=\Gamma_{eq}\frac{L_{c}-L(t)}{x_{B}(t)}$, where
$L_{c}=\sqrt{L_{y}^{2}+\left(L_{x}-x_{0}\right)^{2}}+x_{0}$
($x_{0}=x_B(0)$). In addition, with the above approximations, the
density along the film must also satisfy the local conservation law
$\frac{d\Gamma}{dt}+\Gamma\frac{\partial U}{\partial x}=0$.
Comparison of these two evolution equations leads to $\frac{\partial
U}{\partial x}=\frac{\overset{\centerdot}{x}_{B}}{x_B}\left (1 -
\frac{\Gamma_{eq}}{\Gamma (t)}\cos\alpha (t)\right )$. Note that
since $U\propto x$, the surface velocity at the junction is
$U(x_B(t),t)=\frac{dx_B}{dt}\left (1 - \frac{\Gamma_{eq}}{\Gamma
(t)}\cos\alpha (t)\right )$, which is smaller than the velocity of
the junction itself ($=\overset{\centerdot}{x}_{B}$). This velocity
difference is a consequence of the slip of the surface (and
surfactant) coming from the adjacent film. Finally, Eq.
\ref{equilibrium} is rewritten as an evolution equation for
$x_{B}(t)$:
\begin{multline}
2\gamma_{eq}\left(  \cos\alpha(t)-\frac{1}{2}\right)
+\epsilon\ln\left(
\frac{L_{c}-L(t)}{x_{B}(t)}\right) \label{no diffusion solution}\\
-\left(  \mu_{s}+\kappa\right)  \frac{\overset{\centerdot}{x}_{B}}{x_{B}%
}\left(  1-\frac{x_{B}(t)}{L_{c}-L(t)}\cos\alpha(t)\right)  =0.
\end{multline}

We now compare this theoretical description with the experiments
using measurements of $x_{B}(t)$. From Eq. \ref{no diffusion
solution},  a plot of
$Y(t)=\frac{\overset{\centerdot}{x}_{B}}{x_{B}}\left(  1-\frac{\Gamma_{eq}%
}{\Gamma(t)}\cos\alpha(t)\right)  /\left(
\cos\alpha(t)-\frac{1}{2}\right)  $ versus $X(t)=\ln\left(
\frac{L_{c}-L(t)}{x_{B}(t)}\right)  /\left(
\cos\alpha(t)-\frac{1}{2}\right)  $ should yield a straight line:
$Y=\frac{2\gamma_{eq}}{\mu_{s}+\kappa}+\frac{\epsilon}{\mu_{s}+\kappa}X$.

\begin{figure}[h]
\includegraphics[width=8cm]{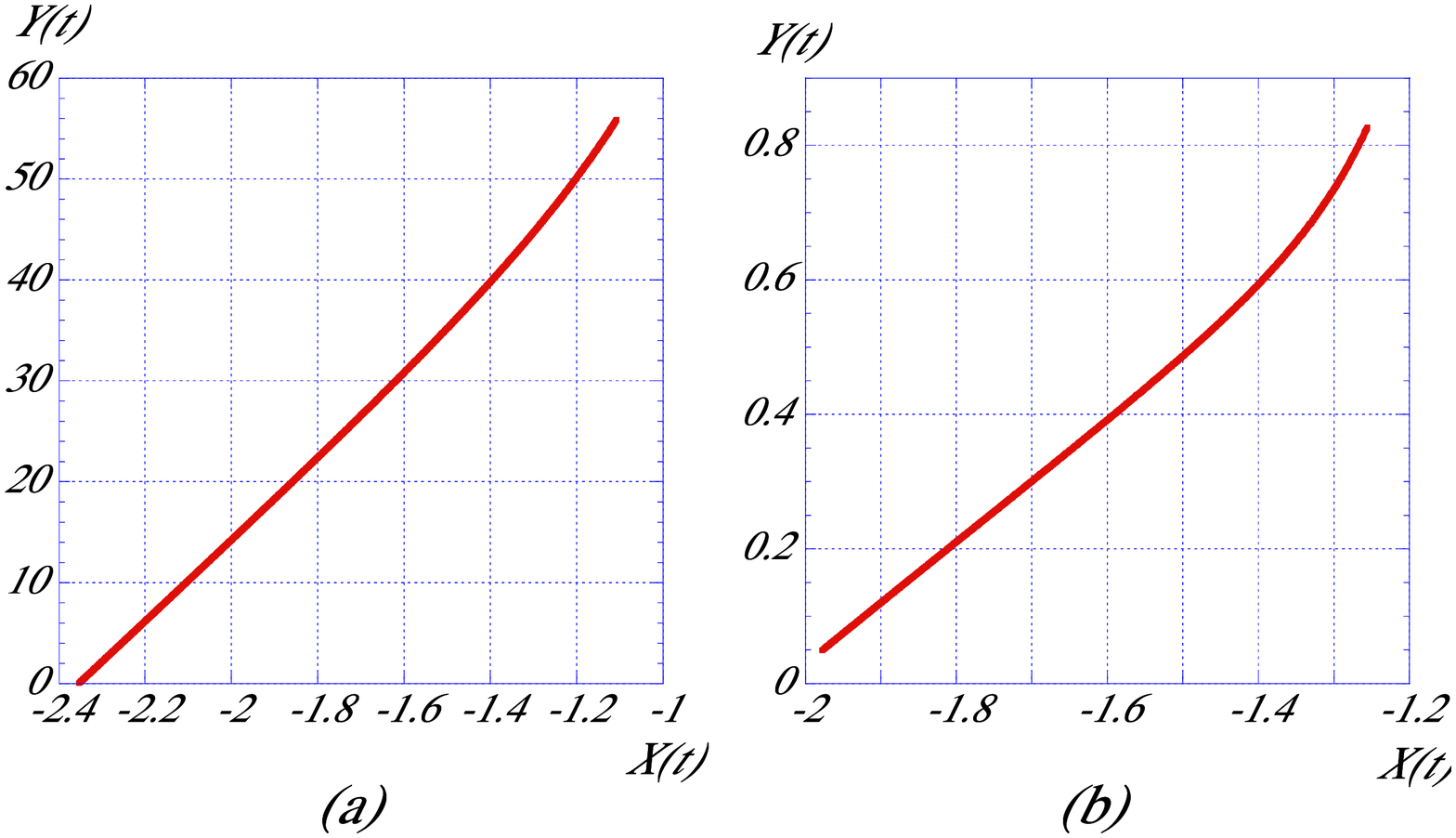}
\caption{(color online) Typical plot of the experimental data (Fig.
\ref{graphs}) reported as $Y(t)=\frac{\overset{\centerdot}{x}_{B}
}{x_{B}}\left( 1-\frac{\Gamma_{eq}}{\Gamma(t)}\cos\alpha(t)\right)
/\left(
\cos\alpha(t)-\frac{1}{2}\right)  $ versus $X(t)=\ln\left(  \frac{L_{c}%
-L(t)}{x_{B}(t)}\right)  /\left(  \cos\alpha(t)-\frac{1}{2}\right)
$; (a) SDS and (b) BSA/PGA. Note the nearly linear evolution in
time, which is in good agreement with the theory. A linear fit of
these curves gives estimates of the surface Gibbs elasticity and the
sum of the
 surface and dilatational viscosities.}
\label{YvsX}
\end{figure}

Typical experimental curves $Y(t)$ vs $X(t)$ are shown in Fig.
\ref{YvsX} for SDS and BSA/PGA \cite{note1}. At short times
(typically, for times below 0.1 sec for SDS and 0.8 sec for BSA),
their evolution is nearly linear, in excellent agreement with the
theory. A linear fit of the experimental data plotted in this way
allows determination of $\mu_s + \kappa$ and $\epsilon$, using
well-established equilibrium surface tension values of SDS and BSA
(38 mN/m \cite{Shen} and 55 mN/m \cite{Graham}, respectively). The
mean values we obtained are $\epsilon=32\pm8$ mN/m and $\mu
_{s}+\kappa=1.3\pm0.7$ mPa.m.s for SDS, and $\epsilon=65\pm12$ mN/m
and $\mu_{s}+\kappa=31\pm12$ mPa.m.s for BSA/PGA, which are in good
agreement with values reported in the literature (e.g.
\cite{Liu,Sarker}). Note, however, that our value of the surface
viscosity of SDS correspond to the highest value reported in
literature \cite{Liu}, which, in both studies, may be a consequence
of the rapid surface stretching giving rise to some nonlinear or
inertial effects. In addition, there is a large difference between
values of the shear and dilatational surface viscosities, and it is
often unclear in published works which surface viscosity is actually
measured.

This study of the T1 dynamics in a two-dimensional foam illustrates
that the relaxation time $T$ associated with the process is a
function of two parameters, $\frac{\mu_s+\kappa}{\gamma_{eq}}$ and
$\frac{\mu_s+\kappa}{\epsilon}$. From dimensional analysis of Eq.
\ref{no diffusion solution},
$T=\frac{\mu_s+\kappa}{\gamma_{eq}}f(\frac{\epsilon}{\gamma_{eq}})$,
where $f$ is an increasing function of the dimensionless parameter
$\frac{\epsilon}{\gamma_{eq}}$. This theoretical description, which
has been corroborated with our experimental data, might be useful
for simulations of aging or rheological properties of foams.
Finally, we expect that a sheared foam has a different rheological
response when the shear rate is significantly different than this
typical relaxation time.

This work was supported by the Harvard MRSEC (DMR-0213805) and
Unilever Research. We thank Legena Jack (and the support of the
Harvard REU program) for assistance with the experiments and Alex
Lips and Jean-Marc di Meglio for helpful conversations.


\begin{thebibliography}{99}
\bibitem{note1}
To obtain $Y(t)$ we need $\overset{\centerdot}{x}_{B}$; first we
fitted $x_{B}(t)$ with a simple function and then took the
derivative. Moreover, for each recorded T1 process, we extract values of $x_{0}$ and $\alpha_{0}$
from the movies, and we account for the asymmetry of the initial configuration by
measuring the different values of $L_x$ and $L_y$ on each side of
the junction.
\end{thebibliography}
\end{document}